\documentclass[a4paper,11pt,twoside]{scrartcl}
\usepackage{ILD}

\usepackage{xspace}

\usepackage{xpatch}
\makeatletter
\xpatchcmd\@HepConStyle
 {\edef\@upcode{\updefault}}
 {\ifdefined\shapedefault\edef\@upcode{\shapedefault}\else\edef\@upcode{\updefault}\fi}
 {}{}
\makeatother

\title{Searches for light exotic scalar decays at the e$^+$e$^-$ Higgs factory}
\ildproc{PHYS}{2026}{007}

\date{February 27, 2026}

\addauthor{Bart{\l}omiej Brudnowski}{\institute{a}}
\addauthor{Kamil Zembaczy{\'n}ski}{} %{\institute{1}}
\addauthor{Aleksander Filip \.Zarnecki}{} %{\institute{1}}

\addinstitute{}{Faculty of Physics, University of Warsaw, Pasteura 5, 02-093 Warsaw, Poland}
\addinstitute{a}{b.brudnowski@student.uw.edu.pl}
\abstract{
The physics program of the Higgs factory will focus on measurements of the 125\,GeV Higgs boson, with the Higgs-strahlung process being the dominant production channel at 250\,GeV. However, similar production of exotic light scalars, in a scalar-strahlug process, is still not excluded by the existing experimental data, provided their coupling to the SM gauge bosons is sufficiently suppressed. This was selected as one of the focus topics of the ECFA Higgs/Top/EW factory study. Presented are the expected cross section limits from the search in the $b\,\bar{b}$ decay channel, based on a full simulation of the International Large Detector (ILD), as well as the expected sensitivity in $\tau^+\tau^-$ and invisible decay channels, based on the fast simulation in the \textsc{Delphes} framework, assuming 250\,GeV ILC running scenario. 

}

\titlecomment{Presented at the International Workshop on Future Linear
  Colliders (LCWS 2025),\\ Valencia, Spain, 20-24 October 2025.\\[5mm]
  This work was carried out in the framework of the ILD Concept Group} %use either this or /onbehalfof above

\graphicspath{{./logos/} {./plots/} }

\begin{document}

\titlepage
\pagenumbering{arabic}\setcounter{page}{2}

\section{Motivation}

% Search for light exotic scalars was selected as one of the so called focus topics for the  ECFA e$^+$e$^-$ Higgs/EW/top factory study~\cite{deBlas:2024bmz}. 
Precision Higgs measurements will be the primary target of a future Higgs factory. At the 250~GeV center of mass energy the SM Higgs production will certainly be the focus, however the existing data does not yet exclude many scalar sector extensions with new scalar states. Light exotic scalars could be copiously produced at a future Higgs factory. However, prospects for new scalar searches were only partially studied in the past. That is why they were selected as one of the focus topics of the ECFA e$^+$e$^-$ Higgs/EW/top factory study~\cite{deBlas:2024bmz}. The primary production channel to be considered is the scalar-strahlung process, $e^+ e^- \to Z \; S $, the light scalar analogue of the Higgs-strahlung process, $e^+ e^- \to Z \; H $, which is the dominant production process for the SM Higgs boson at 240--250~GeV. 

The studies presented below were carried out in the framework of the ILD concept group and based on the assumed ILC running scenario \cite{Barklow:2015tja}, but the results should be applicable also to other e$^+$e$^-$ Higgs factories operating in this energy range.

% \section{ILC and its experiments}

\section{Analysis framework}

Exotic scalar production at the ILD experiment was previously studied mainly in the decay-mode independent approach \cite{Wang:2018awp,Wang:2020lkq}.
The advantage of it is that no assumptions are made concerning the new scalar decay channel and the identification is purely based on the scalar mass reconstruction from the four-momentum conservation (so called recoil mass).
However, much higher sensitivity can be reached, if particular decay channel of the new scalar is assumed.
Considered in this contribution are scalar decays to $b\bar{b}$, $\tau^+\tau^-$ as well as invisible decays.

\subsection{Full simulation for the  $S \to b\,\bar{b}$ channel}

As the $S \to b\bar{b}$ decay is the dominant decay channel for a wide mass range of masses, for SM-like light scalars, the study addressing this decay could be based on the Monte Carlo samples previously generated for the decay mode independent analysis.

For the background, existing ILD full simulation Monte Carlo sample for the SM processes at 250~GeV were used \cite{Berggren:2021sju}. The samples were generated using \textsc{Whizard} \cite{Kilian:2007gr} v.2.8.5 using Set A ILC beam-spectrum and ILD\_l5\_o2\_v02 model, simulated and reconstructed in \textsc{Geant4} \cite{GEANT4:2002zbu} and \textsc{ILCSoft} \cite{Ete:2021ljr} v02-02-01. Jet clustering and flavour tagging were performed using \textsc{Marlin} \cite{Gaede:2006pj} and \textsc{LCFIPlus} package \cite{Suehara:2015ura}.

Signal was generated and and reconstructed with the same tools but the detector simulation was done using \textsc{SGV} \cite{Berggren:2012ar}. Signal samples were generated for masses ranging from 10~GeV to 160~GeV with a step of 5~GeV. The H-20 running scenario was assumed with the total integrated luminosity of 2~ab$^{-1}$, $\pm30\%$ positron and $\pm80\%$ electron beam polarisation \cite{Barklow:2015tja}.

\subsection{Fast simulation for $S\to \tau^+\tau^-$ and $S\to$\emph{invisible} channels}

For $S\to \tau^+\tau^-$ and $S\to$\emph{invisible} channels the analysis was based on the fast simulation of the detector response. Dedicated signal and background samples were generated with \textsc{Whizard} 3.1.2 using the built-in SM\_CKM model. Signal was generated by varying the mass of the Higgs boson and forcing its decay to the considered final state which was $\tau^+\tau^-$ or $Z\,Z^{(*)} \to \nu\bar{\nu}\,\nu\bar{\nu}$ for the invisible decay channel. Only hadronic Z decays were considered.

For the background, all relevent four-fermion final states were considered as well as quark pair-production, while six-fermion final states were neglected due to their small cross-section at $\sqrt{s}=$250\,GeV. 
For invisible scalar decays, processes involving photons in the initial state, $\gamma e^\pm$ and $\gamma \gamma$ (beamstrahlung photons as well as collinear photons described and Equivalent Photon Approximation, EPA) were also taken into account. The H-20 scenerio for ILC running was assumed as before. The detector simulation and event reconstruction was done in \textsc{Delphes} \cite{deFavereau:2013fsa} assuming ILCgen detector model.

\section{Analysis results}

\subsection{$S \to b\, \bar{b}$}
%  Assuming SM-like light scalar couplings, $bb$ channel would be dominating for lower masses of exotic scalar particles. Here the analysis was carried out in the framework of the ILC for masses between 10 and 160\,GeV for the H-20 running scenario.

\begin{figure}[t]
    \centering
    \includegraphics[width=0.48\linewidth]{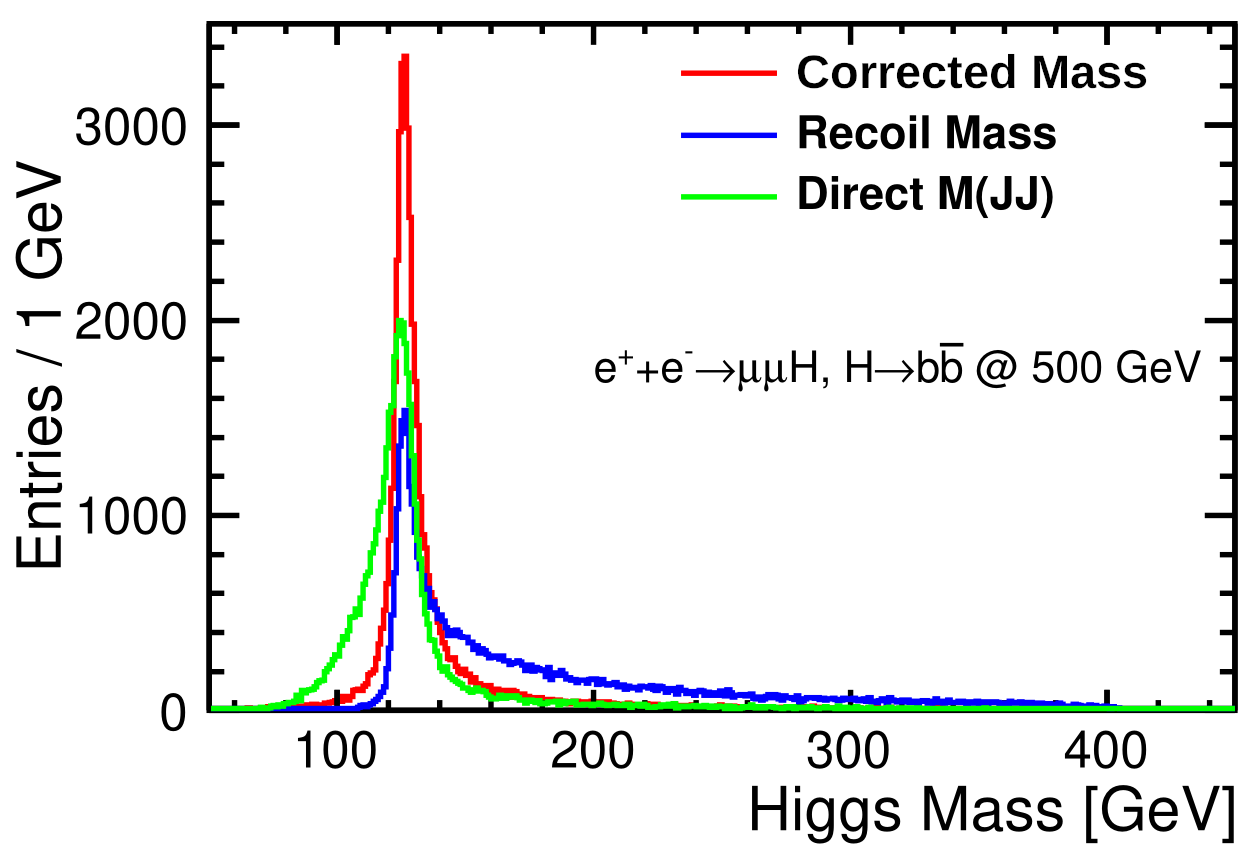}%
    \includegraphics[width=0.48\linewidth, trim=0 0 1cm 1cm, clip]{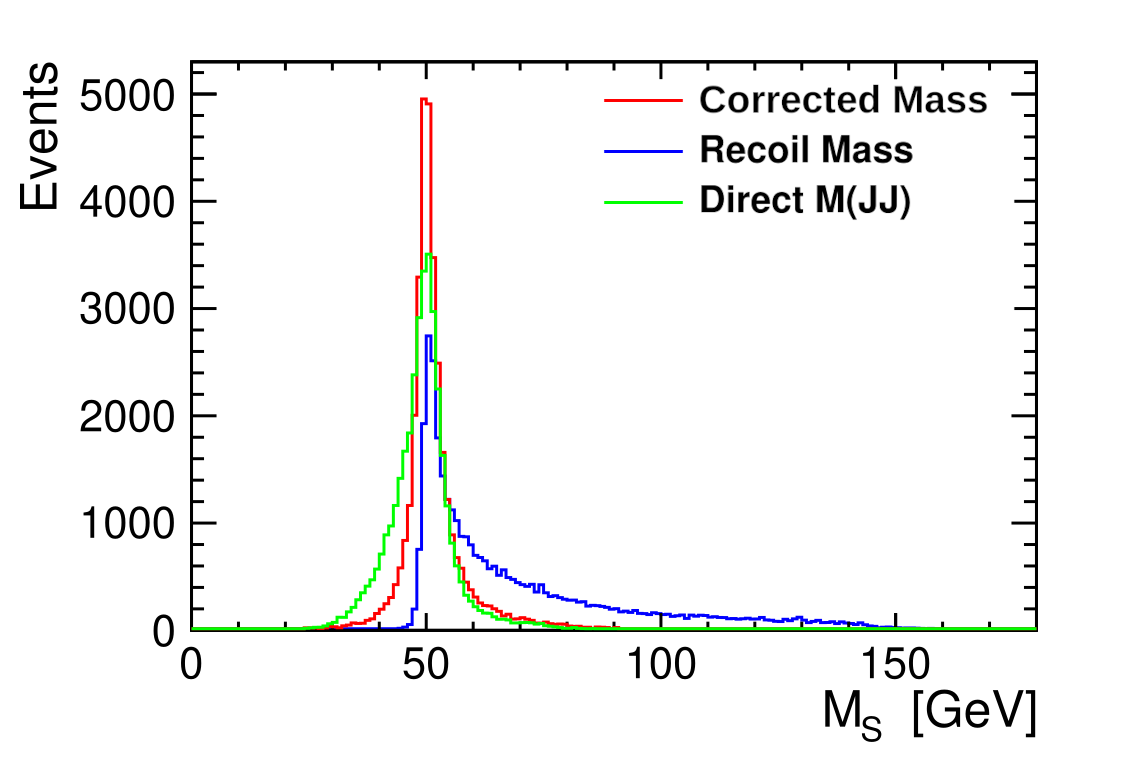}
    \caption{Left: Invariant mass of a reconstructed Higgs boson from a full simulation of 500\;GeV run sample using three methods: direct reconstruction from measurement momenta, recoil mass and reconstruction from corrected momenta. Figure from \cite{Tian:2020,Berger:2024pwj}.
    Right: Invariant mass of a reconstructed scalar of 50\;GeV mass from a fast simulation of 250\;GeV run sample using the same three methods: direct reconstruction from measurement momenta, recoil mass and reconstruction from corrected momenta.}
    \label{fig:higgs_reco}
\end{figure}

Presented here are results from the study focused on the leptonic Z boson decays, $Z\to e^+e^-$ and $Z\to \mu^+\mu^-$. The analysis also covering the hadronic Z decay channel is still ongoing.

Event pre-selection stage involved reconstructing the light scalar from its hadronic decay products (two $b$ jets) and the Z boson from the lepton pair. Since jet energy reconstruction is much less precise, a correction method based on the measured momentum of the lepton pair was applied to increase the resolution \cite{Tian:2020,Berger:2024pwj}. Compared in Fig.~\ref{fig:higgs_reco} (left) is the distribution of the SM Higgs boson mass reconstructed with three methods: direct reconstruction from the pair of $b$ jets, recoil mass reconstruction and the mass reconstructed after jet momenta correction. As shown in Fig.~\ref{fig:higgs_reco} (right), the proposed correction works very well also for light scalar production and was therefore used in the presented study. 

\begin{figure}
    \centering
    \includegraphics[height=4.4cm, trim=6.7cm 5.6cm 0 0.5cm, clip]{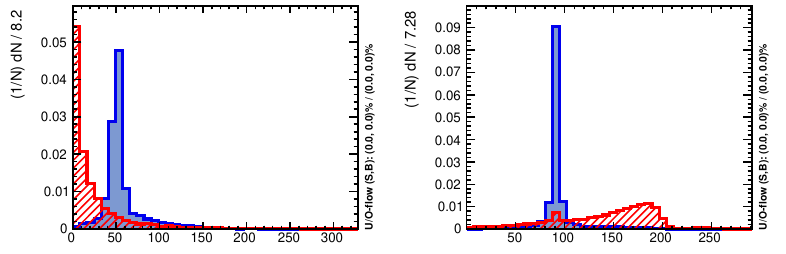}
    \hspace*{4cm}\tiny\textbf{M$_S$ [GeV]}
    \hspace*{5.5cm}\textbf{M$_Z$ [GeV]}
    \caption{Normalised distributions of the reconstructed invariant mass for SM background events (red) and signal of 50\,GeV scalar production (blue), for LR beam polarisation settings. Left: invariant mass for the two reconstructed hadronic jets, after jet momentum correction. Right: invariant mass of the lepton pair.}
    \label{fig:invmass_dist}
\end{figure}

Using the momentum correction method introduced above, invariant mass of the scalar particle could be calculated from corrected jet four-momenta. Distributions of the reconstructed scalar candidate mass for SM background and 50\,GeV signal scenario are compared in Fig.~\ref{fig:invmass_dist}, together with the corresponding distributions for the Z candidate (lepton pair). 
Very good separation of signal and background events is clearly visible in both cases.

\begin{figure}
    \centering
    \includegraphics[height=4.7cm, trim=6.7cm 0 0 5.7cm, clip]{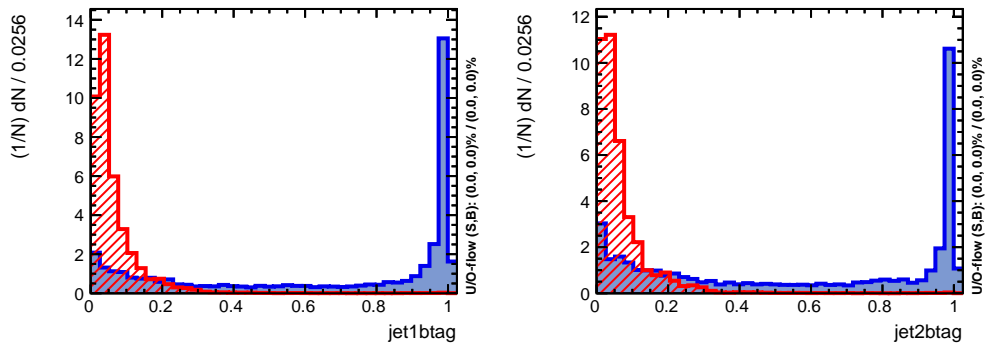}
    \caption{Normalised distribution of the LCFIPlus b-tagging algorithm response for the two reconstructed hadronic jets  for SM background events (red) and signal of 50\,GeV scalar production (blue), for LR beam polarisation setting.}
    \label{fig:btagging_dist}
\end{figure}

Flavour tagging also turned out to be crucial  for the efficient background suppression in the presented study. 
% A crucial variable for the classification was the b-tagging response. 
Jet clustering and flavour tagging was carried out using the LCFIPlus package. Distributions of the b-tagging response for the two reconstructed jets are presented in Fig.~\ref{fig:btagging_dist}. 
Also on the b-tagging level itself, we observe clear separation between signal and the background processes.

\begin{figure}
	\centering
	\includegraphics[width=.48\linewidth, trim=0 0 0 1.2cm, clip]{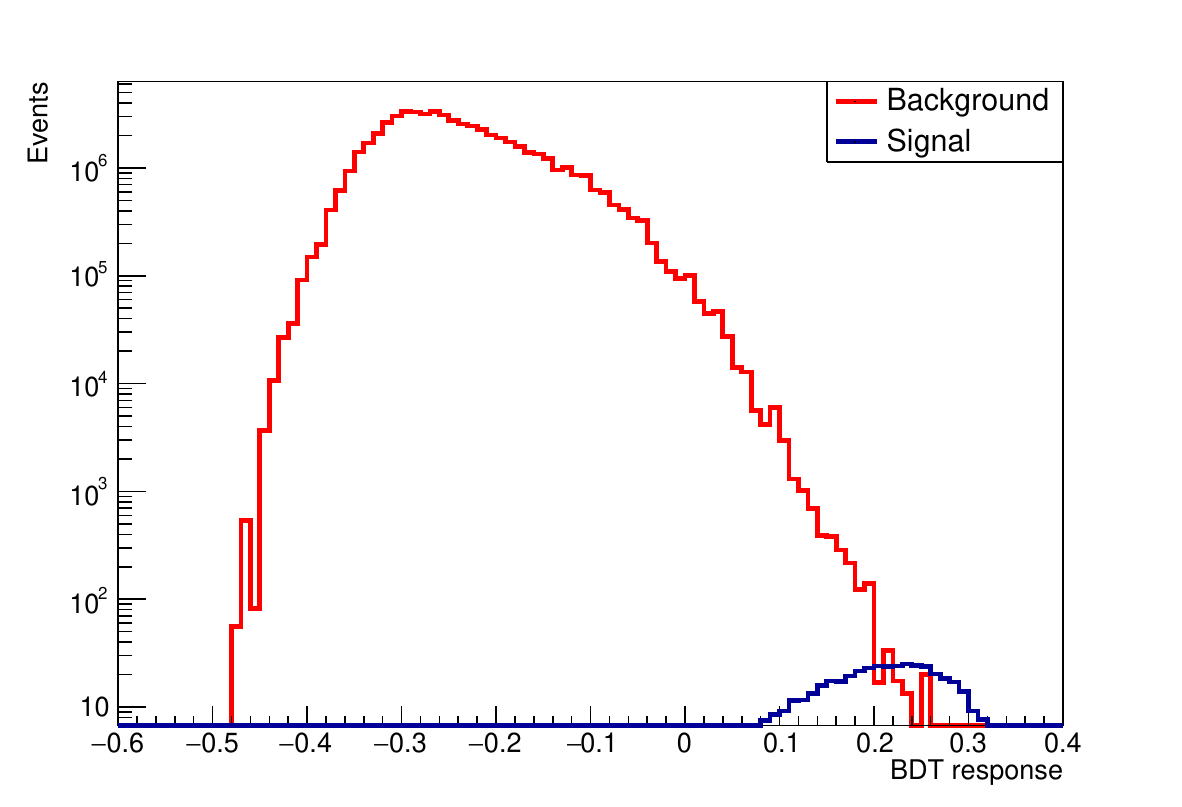}%
	\includegraphics[width=.48\linewidth, trim=0 0.7cm 0 0, clip]{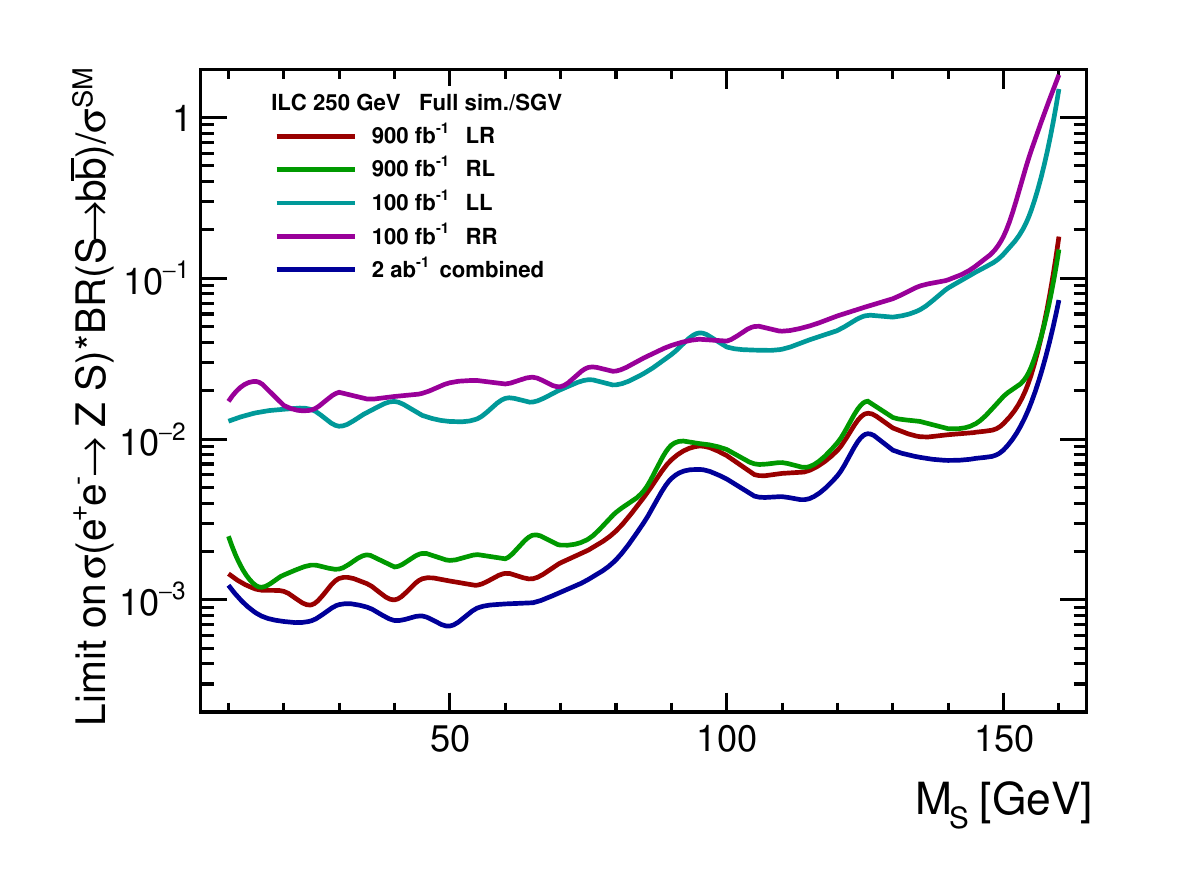}%
	\caption{Left: BDT response distribution for 50~GeV signal (blue) and SM backgrounds (red) for the LR polarisation running at 250\,GeV. Signal was scaled to 1\% of the SM Higgs production cross section. Right: Expected limits for the exotic scalar production times the $b\bar{b}$ branching ratio for scalar mass from 10 to 160\,GeV. Separate limits for each beam polarisation setting are plotted as well as the combined limit.}
	\label{fig:bdt_limit_bbll}
\end{figure}

A Boosted Decision Tree (BDT) classifier based on TMVA \cite{TMVA:2007ngy} was trained using kinematic variables of the event and the b-tagging responses. A classifier was trained separately for a combination of 31 different masses of the scalar and 4 different beam polarisation settings considered in H-20 running scenario. Using the shape of the BDT response histogram for signal and background, as seen in the Fig.~\ref{fig:bdt_limit_bbll} (left) the 95\% C.L limits on the scalar production cross section could be extracted. 
The limits were first calculated for each scalar particle mass and different beam polarisation setting. Then a combined limit was obtained for each mass including all beam polarisation settings. 
Results are compared in Fig.~\ref{fig:bdt_limit_bbll} (right). 
For a given mass of the new scalar they are shown
as the excluded cross section relative to the production cross section the SM Higgs boson would have had if its mass was that of the new scalar.
For low scalar masses combined limits go down to below $10^{-3}$ of the SM cross section. Search sensitivity decreases significantly for scalar masses approaching 90~GeV and 125~GeV reflecting Z-pair production and SM Higgs production processes becoming dominant backgrounds in this mass range.

\subsection{$S \to \tau^+ \tau^-$}

Analysis of the $S \to \tau^+ \tau^-$ decay channel, while  performed on the fast simulation level, follows the similar procedure as described above.  We focus on hadronic Z decays in this case, as the signal cross section is an order of magnitude higher than for the leptonic Z decay due to the higher Z decay branching ratio and, similar to the choice of leptonic Z decays for $S \to b\bar{b}$, this allows us to avoid ambiguities in event reconstruction.

Three event categories can be considered depending on the assumed decays of the two $\tau$ leptons from the decay of the new scalar:
hadronic (with both $\tau$ decaying hadronically), semi-leptonic (with one leptonic $\tau$ decay) and leptonic (with leptonic decays of both $\tau$).
Two selection approaches were also considered: tight selection, when we require two identified $\tau$ lepton candidates in an event (either an isolated lepton or hadronic jet with $\tau$-tag) and the loose selection when we require only one $\tau$ candidate and take the untagged jet with the smallest invariant mass as the second one.

\begin{figure}
	\centering
	\includegraphics[width=.48\linewidth]{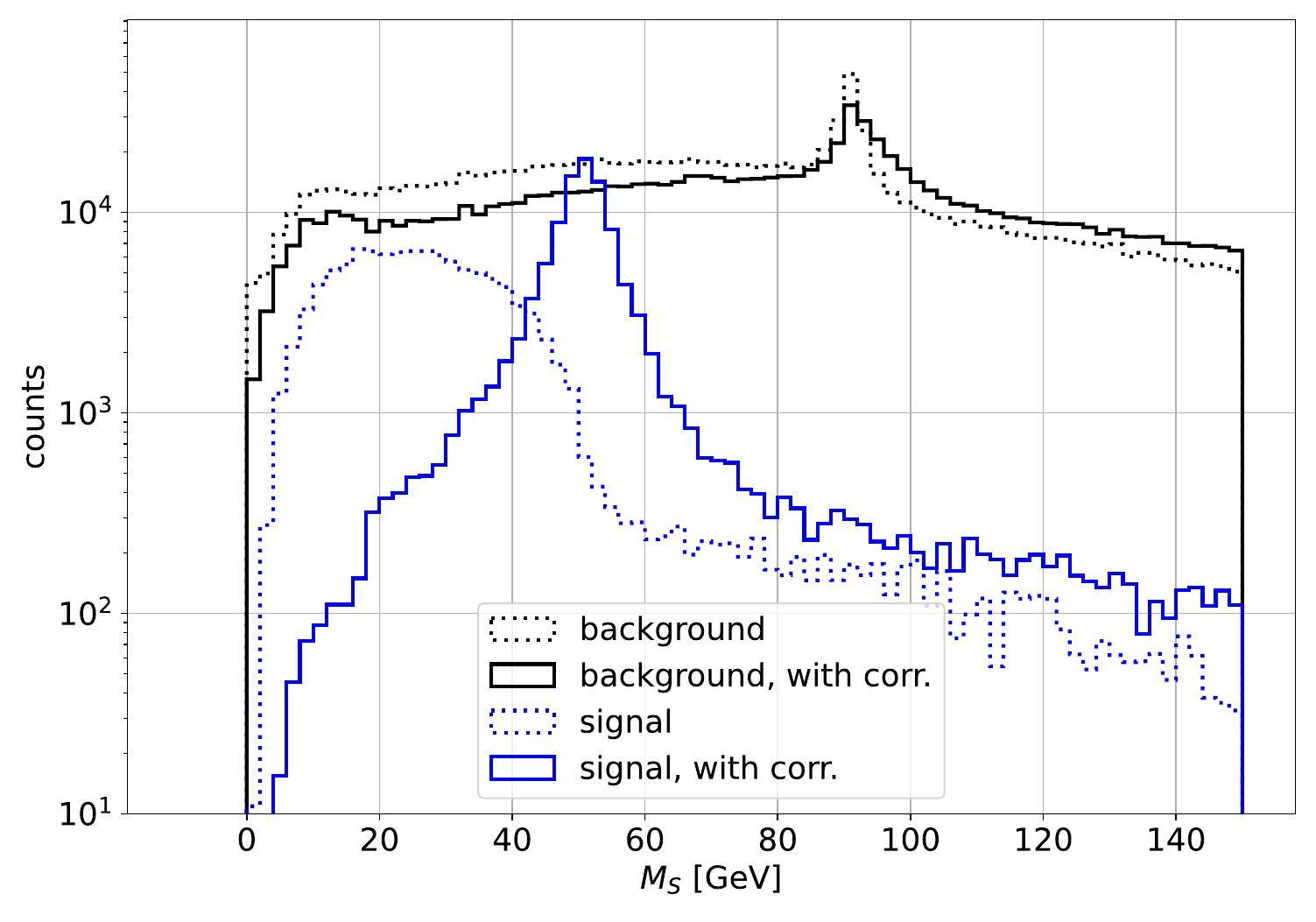}%
	\includegraphics[width=.48\linewidth]{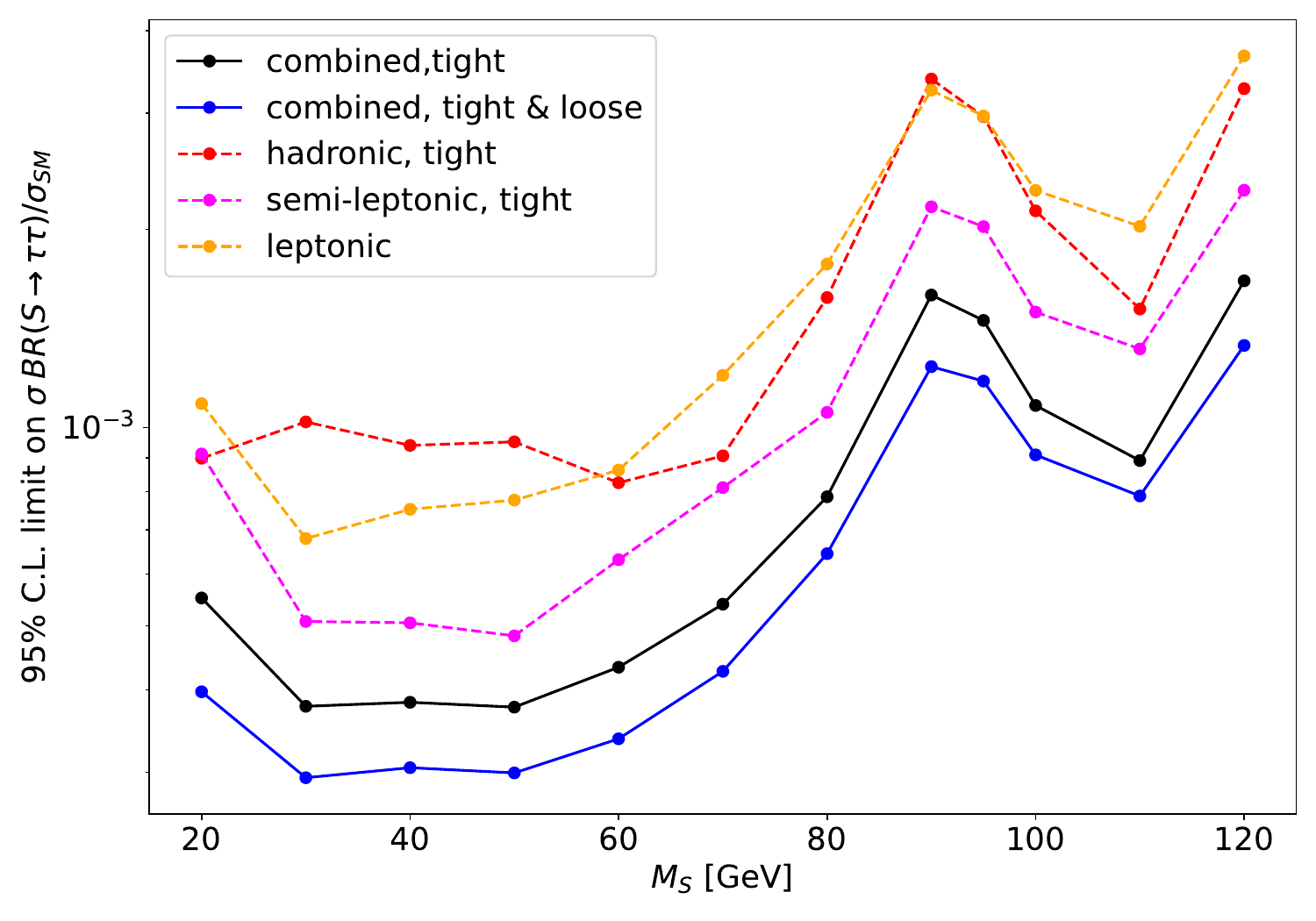}%
	\caption{Left: Reconstructed invariant mass of the $\tau$ lepton pair, before and after energy correction, for SM background processes and signal of 50\,GeV scalar production, LR polarisation settings and tight selection of semi-leptonic events. Signal was scaled to 1\% of the SM Higgs production cross section for mass of 50\,GeV. Right: Expected limits for exotic scalar production times the $\tau^+\tau^-$ decay branching ratio for scalar mass from 20 to 120\,GeV. Presented are limits obtained for different event categories and the combined limit. Results contributed to \cite{Altmann:2025feg}.}
	\label{fig:tautau}
\end{figure}

For the scalar invariant mass reconstruction, a correction based on the collinear approximation is applied \cite{Kawada:2015wea} to the $\tau$ candidate energy, to correct for the missing neutrino momentum. 
An example of the correction performance for the semi-leptonic event sample after tight selection is presented in Fig.~\ref{fig:tautau} (left). For the signal of 50\,GeV scalar production a sharp peak in the invariant mass distribution is nicely reconstructed at the nominal mass when the correction is applied.
After event classification and energy correction, BDT classifiers were trained for different beam polarisations and event categories, and the combined limits were then calculated.
Results are presented in Fig.~\ref{fig:tautau} (right).
The semi-leptonic event category turned out to be most sensitive to the light scalar production, combining high event statistics (about 47\% of decays) with lower background than for the hadronic channel.  
The combined limits based on tight event selection are further improved by 20-30\% by including the loose selection categories.

\subsection{$S \to $\emph{invisible}}

\begin{figure}
	\centering
	\includegraphics[width=.55\linewidth]{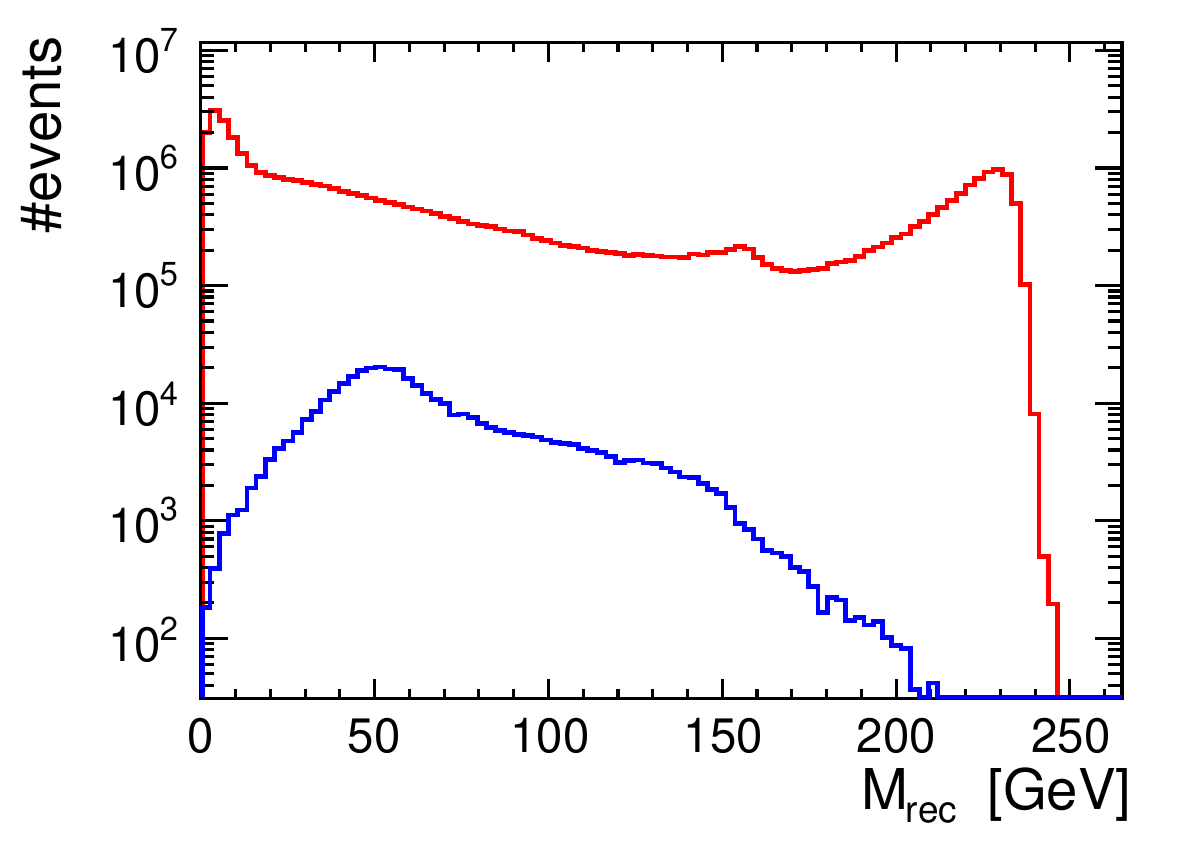}%
	\includegraphics[width=.43\linewidth,trim=0 -1cm 0 0, clip]{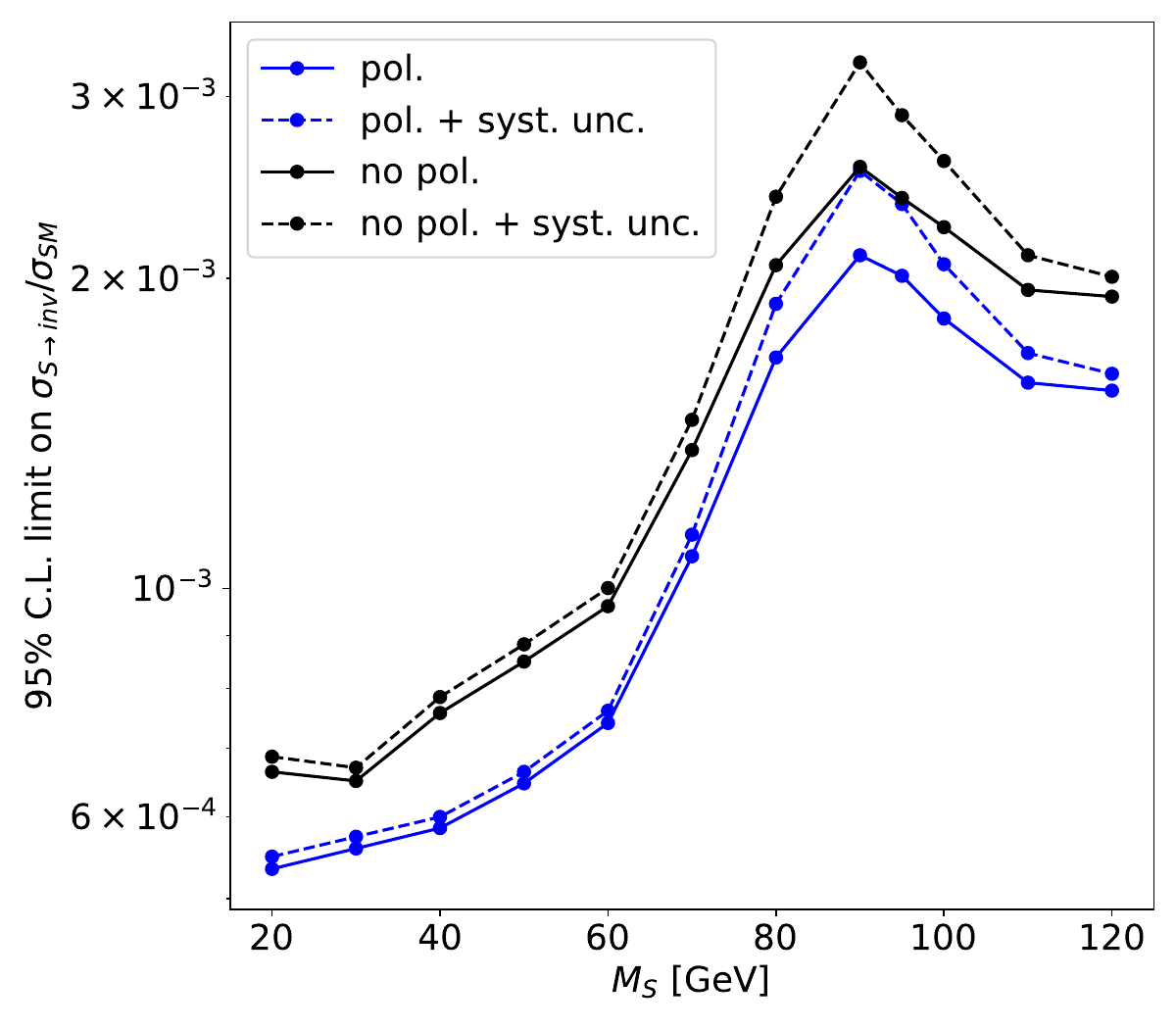}%
	\caption{Left: Reconstructed recoil mass for SM background processes (red) and signal of 50\,GeV scalar production with invisible scalar decays (blue). Signal was scaled to 1\% of the SM Higgs production cross section for mass of 50\,GeV. Right: Expected limits for exotic scalar production times the invisible branching ratio, relative to the SM Higgs production cross section, for scalar mass from 20 to 120\,GeV. Compared are limits obtained for polarised beams (assuming H-20 luminosity sharing scenario) and for unpolarised beams, without and with systematic uncertainties taken into account. Results contributed to \cite{Altmann:2025feg}.}
	\label{fig:invisible}
\end{figure}

For the invisible decay channel, we looked for events with a single Z reconstructed in the  hadronic decay channel and no other activity in the detector. 
An additional preselection based on the invariant mass of the reconstructed Z boson (required to fit between 74\,GeV and 114\,GeV) and its transverse momentum (above 10\,GeV) was applied to suppress SM backgrounds.  
The signal of light scalar production should be visible in the reconstructed recoil mass distribution, as shown in Fig.~\ref{fig:invisible}.
A BDT classifier was then trained with parameters of the reconstructed Z boson as input variables for final signal-background discrimination. Limits resulting from the obtained BDT response distributions are presented in Fig.~\ref{fig:invisible} (right). 
Compared are limits obtained for polarised and  unpolarised beams. 
The combined analysis of the four polarisation configurations, as included in the H-20 running scenario,
results in about a 20\% improvement with respect to the same integrated luminosity with unpolarised beams.
Also shown is the impact of the systemic uncertainties related to normalisation of the theory predictions and normalisation of the data samples --- it turns out to be significant only in the scalar mass range around the $W$ and $Z$ boson masses.

\section{Summary and Conclusions}

\begin{figure}
	\centering
	\includegraphics[width=.8\linewidth]{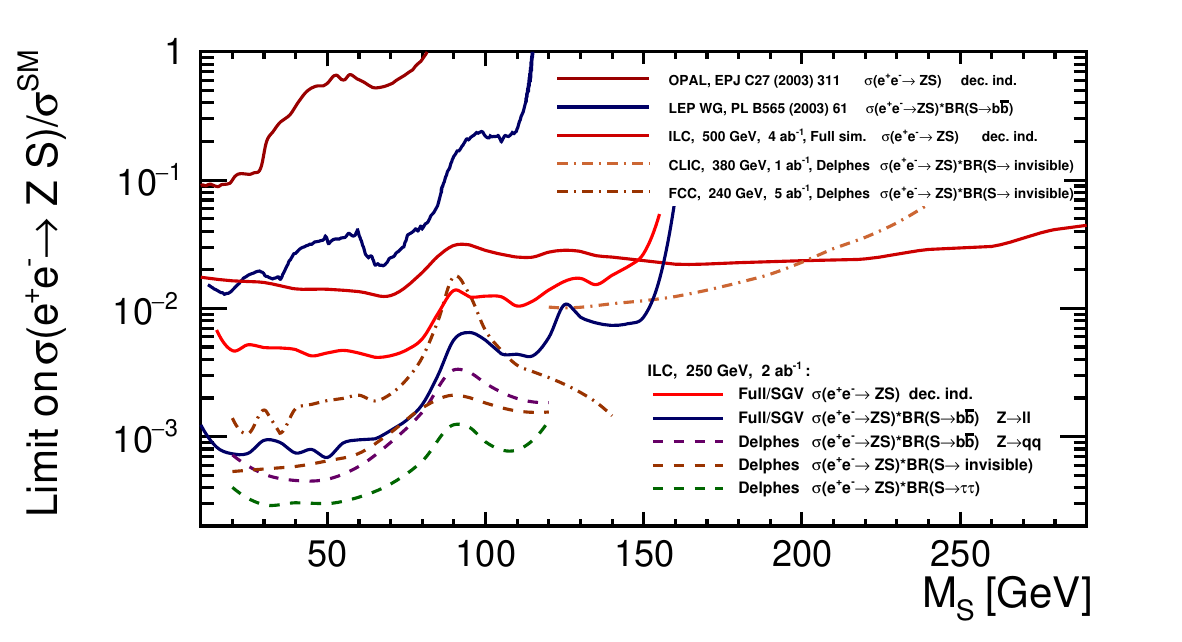}
	\caption{Comparison of expected exclusion limits for exotic scalar production searches in the scalar-strahlung processes, for different scalar decay channels.  Figure taken from \cite{Altmann:2025feg}.}
	\label{fig:summary}
\end{figure}

Light scalar states which could be produced at e$^+$e$^-$ Higgs factories are still not excluded by the experimental data. Experiments at future e$^+$e$^-$ colliders can probe this sector and set stringent limits in multiple channels. Comparison of expected exclusion limits is shown in Fig.~\ref{fig:summary}. While the decay mode independent searches are sensitive to cross sections down to $4\cdot 10^{-3}$ of the SM production cross section, even higher sensitivity can be reached when focusing on particular decay channels. For the $b\bar{b}$, dominant decay channel for light scalars in many models, limits on the branching ratio times the cross section ratio go down to below $10^{-3}$. These limits are likely to improve further when including hadronic Z decay channels. The highest search sensitivity is expected in the $\tau$ decay channel, provided the scalar branching ratio for this channel is significant.

\subsection*{Acknowledgments}

This work was supported by the National Science Centre,
Poland, under the OPUS research project no. 2021/43/B/ST2/01778.

\printbibliography

\end{document}